\documentclass[conference,10pt]{IEEEtran}
\IEEEoverridecommandlockouts
\def\BibTeX{{\rm B\kern-.05em{\sc i\kern-.025em b}\kern-.08em
    T\kern-.1667em\lower.7ex\hbox{E}\kern-.125emX}}

\usepackage[utf8]{inputenc}
\usepackage{cite}
\usepackage{amsmath,amssymb,amsfonts}
\usepackage{algorithmic}
\usepackage{graphicx}
\usepackage{textcomp}
\usepackage{subcaption}
\usepackage[ruled,vlined]{algorithm2e}
\usepackage{epsfig}
\usepackage{float}
\usepackage{color}
\usepackage{balance}
\usepackage{bbm}
\usepackage{textcomp}
\usepackage{enumitem}
\usepackage{comment}

\usepackage{mathtools,nccmath,amsthm}
\usepackage{multicol,tikz,array,booktabs,url}
\usepackage{xcolor}
\usepackage{acronym}
\usepackage{xfrac}

\graphicspath{{figures/}{./}}
\DeclareGraphicsExtensions{.pdf}

\DeclareMathAlphabet{\mathpzc}{OT1}{pzc}{m}{it}

\begin{document}
\title{Near-Field Beam Focusing Characterization\\ for 2D Waveguide-Fed Metasurface Antennas}

\author{Panagiotis Gavriilidis$^1$ and George C. Alexandropoulos$^{1,2}$\\
$^1$Department of Informatics and Telecommunications, National and Kapodistrian University of Athens, Greece
\\$^2$Department of Electrical and Computer Engineering, University of Illinois Chicago, USA
\\e-mails: \{pangavr, alexandg\}@di.uoa.gr
}

\maketitle
\thispagestyle{empty}

\begin{abstract}
Two-dimensional (2D) waveguide-fed metasurfaces enable scalable antenna apertures through guided wave excitation of distributed radiating elements. However, the resulting non-uniform excitation challenges classical interpretations of near-field characteristics. Using a physics-compliant model, this paper analyzes the near-field beam focusing behavior of such architectures. We derive asymptotic scaling laws for the beamforming gain, showcasing that the power-normalized gain scales linearly with the number of radiating elements. Furthermore, we introduce a normalized beam-depth formulation and obtain a compact analytic expression that characterizes the transition to far-field-like behavior. The presented analysis is validated against simulations based on the full electromagnetic model, confirming the accuracy of the derived scaling laws and beam-depth limits.
\end{abstract}

\vspace{-0.4 cm}
\section{Introduction}
Metasurface-based antennas are envisioned as a key technology for future Sixth-Generation (6G) wireless networks, enabling low-cost and power-efficient implementations of eXtremely Large (XL) antenna arrays with increased spatial degrees of freedom and gain. Due to their large physical size, the radiative Near-Field (NF) region is significantly extended, requiring accurate modeling of wavefront curvature.  Conventional criteria for the transition between Far-Field (FF) and NF, such as the Rayleigh distance, are based on phase error arguments. However, a more meaningful definition relates to beamforming capability, i.e., identifying the region where spherical wavefronts must be accounted for in beam design~\cite{Bjornson_dist}. In this context, the NF beam focusing behavior and, in particular, the depth of focus for planar arrays, was introduced in~\cite{Bjornson_dist} for targets located along the array normal. Subsequently,~\cite{SDMA_vs_LDMA} derived analytical approximations of beam orthogonality and proposed spherical-domain beamforming codebooks.


However, the prescribed works focus on ideal apertures with uniform excitation, which does not hold for metasurface-based antennas. In these systems, the metamaterials are excited by guided waves, where propagation, dielectric, and radiative losses induce a non-uniform excitation profile across the aperture. The NF focusing behavior of such architectures was first investigated in \cite{gavriilidis2025LossyDMA} and, specifically, for Dynamic Metasurface Antennas (DMAs) implemented using stacked microstrip lines with reconfigurable metamaterials, which were modeled via lossy transmission-line theory. In contrast, this paper considers two-Dimensional (2D) waveguide-fed DMAs \cite{pulidomancera2018}, which provide a more natural platform for realizing XL antenna apertures, since the entire planar structure can be excited by a single feed and then scaled by enlarging the Parallel Plate Waveguide (PPW), without requiring additional Radio-Frequency (RF) circuitry. Moreover, we adopt a full electromagnetic model that accounts for mutual coupling and dual-polarized interactions, and derive an analytical expression for the NF beam depth, along with gain scaling laws with respect to (w.r.t.) the number of metamaterial elements.

\section{System and Channel Modeling}
We consider a Transmitter (TX) implemented as a metasurface-based antenna realized through a PPW of height \(h\), filled with vacuum \cite{pulidomancera2018}. The metasurface is excited by a thin wire feed placed inside the waveguide, and radiates through subwavelength reconfigurable metamaterial elements etched on the upper plate. Each element resembles a polarizable dipole, and the whole metasurface can be described using the coupled dipole formalism~\cite{pulidomancera2018}. 
Specifically, each \(n\)-th (\(n=1,\ldots,N\)) metamaterial is characterized by a polarizability matrix \(\mathbf{A}_n \triangleq [A^{xx}_{n},A^{xy}_{n}; A^{yy}_{n}, A^{yx}_{n}]\) that relates its local magnetic field \(\mathbf{h}_{\text{loc},n}\in\mathbb{C}^{2\times 1}\) to the induced magnetic moment \(\mathbf{m}_n\triangleq[m^x_{n},m^y_{n}]^{\rm T}\). We have assumed the metasurface lies on the \(x\)--\(y\) plane at \(z=0\), so only the \(x\) and \(y\) magnetic components are relevant, since the \(z\) component of the magnetic field is zero. The relation between \(\mathbf{m}_n\) and \(\mathbf{A}_n\)~is:
\begin{equation}\label{eq: magnetic moment from local field}
    \mathbf{m}_n = \mathbf{A}_n \mathbf{h}_{\text{loc},n}.
\end{equation}

The local magnetic field can be expressed as the sum of the field from the source $\mathbf{h}_{0,n}=[h^{x}_{0,n},h^y_{0,n}]\in \mathbb{C}^{2 \times 1}$ and the scattered fields generated by all other metamaterials, yielding:
\vspace{-0.2 cm}
\begin{equation}
    \mathbf{h}_{\text{loc},n} = \mathbf{h}_{0,n} + 
    \sum_{j=1,j\neq n}^{N}\mathbf{G}_{n,j}\mathbf{m}_j,
    \label{eq:Hloc_expanded_final}
\end{equation}
where $\mathbf{G}_{n,j}\in \mathbb{C}^{2 \times 2}$ models the coupling between the $j$-th and $n$-th elements. Letting $\mathbf{r}_n \triangleq [x_n,y_n,0]$ denote the position of the $n$-th element, the interactions are modeled as follows:
\begin{equation}\label{eq: Total Green Function}
\mathbf{G}_{n,j} = 
\begin{cases}
\mathbf{G}_{\text{WG}}\left(\mathbf{r}_n-\mathbf{r}_j\right)+ \mathbf{G}_{\text{FS}}\left(\mathbf{r}_n-\mathbf{r}_j\right), & n\neq j\\
0, & n=j
\end{cases},
\end{equation}
with $\mathbf{G}_{\text{WG}}$ describing the coupling through the waveguide and $\mathbf{G}_{\text{FS}}$ the coupling through free space for \(x\) and \(y\) polarized fields. For the considered case of a PPW, the waveguide and free-space contributions are given in \cite[eqs.~(4) and (5)]{gavriilidis2026nearfield}. In this paper, we study the case where a single feed exists, implying that the beamforming capabilities of the PPW-fed metasurface will be isolated, and the beam focusing limits can be studied. Furthermore, we consider the feed to be a current source with current denoted by \(I\), which is placed at the center of the PPW, i.e., \(\mathbf{p}\triangleq [0,0,0]\); for this case, the magnetic field induced in each \(n\)-th element due to the feed can be expressed as~\cite{pulidomancera2018} ($H^{(2)}_{1}(\cdot)$ is the $2$nd kind, $1$st order $\nu$ Hankel function):  
\begin{equation}\label{eq: Excitation Field}
\begin{split}
    h^{x}_{0,n}\!\!=& \frac{\jmath k}{4}I \, H^{(2)}_{1}\!\!\left(k\rho_n\right)\sin\!\left(\psi_n\right),\\
    h^{y}_{0,n}\!\!=& \frac{\text{-}\jmath k}{4} I \, H^{(2)}_{1}\!\!\left(k\rho_n\right)\!\cos\!\left(\psi_n\right),
\end{split}
\end{equation}
where we have used the definitions \(\rho_n \triangleq \sqrt{x^2_n + y^2_n}\) and \(\psi_n \triangleq \mathrm{atan}\!\left( (p_{y}- {y_n})/(p_{x}- {x_n})\!\right)\).

Let us define the concatenated dipole moment vector $\mathbf{m} \triangleq [\mathbf{m}_1,\ldots,\mathbf{m}_N]  \in \mathbb{C}^{2N\times 1}$, the polarizability matrix $\bar{\mathbf{A}} \triangleq {\rm diag} [\mathbf{A}_1,\ldots,\mathbf{A}_N] \in \mathbb{C}^{2N\times 2N}$ containing the polarizabilities of all elements, i.e., \([\bar{\mathbf{A}}]_{2n-1:2n,2n-1:2n}=\mathbf{A}_n\) $\forall n$, and the mutual coupling array \(\bar{\mathbf{G}}\in \mathbb{C}^{2 N \times 2 N}\) for which \([\bar{\mathbf{G}}]_{2n-1:2n,2j-1:2j}=\mathbf{G}_{n,j}\) \(\forall \,n,j=1,\ldots,N\). Additionally, the excitation-field vector $\mathbf{h}_0  \in \mathbb{C}^{2N\times 1}$ is given by \(\mathbf{h}_0=\mathbf{h}_{\rm f} I \), where we distinguish between the propagation functions \(\mathbf{h}_{\rm f}\) and the current \(I\), with \([\mathbf{h}_f]_{2n-1}=\frac{\jmath k}{4}H^{(2)}_{1}\!\left(k\rho_n\right)\sin\!\left(\psi_n\right)\) and \([\mathbf{h}_f]_{2n}=\frac{-\jmath k}{4}H^{(2)}_{1}\!\left(k\rho_n\right)\cos\!\left(\psi_n\right)\). Using these definitions, and since \eqref{eq: magnetic moment from local field} and \eqref{eq:Hloc_expanded_final} hold for all elements simultaneously, by solving w.r.t. \(\mathbf{m}\), we obtain:\vspace{-0.1cm}
\begin{equation}\label{eq: Dipole Moment from currents}
    \mathbf{m} = \left(\bar{\mathbf{A}}^{-1} - \bar{\mathbf{G}}\right)^{-1}\mathbf{h}_f I.\vspace{-0.1cm}
\end{equation}
This expression fully characterizes the coupled dipole formalism. Furthermore, to model the responses of the metamaterials, we use the Radiation-Reaction (RR) corrections, which ensure passivity for the elements. Specifically, the RR corrections tailored for the PPW architecture are given in~\cite{Gavriilidis2026WaveguideMetasurface} as follows: 
\vspace{-0.1cm}
\begin{equation}\label{eq: rr_correction}
\mathbf{A}_n
= \tilde{\mathbf{A}}_n \left(\mathbf{I}_2 + \jmath C\tilde{\mathbf{A}}_n\right)^{-1},\vspace{-0.1cm}
\end{equation}
where \(C \triangleq (k^3/(3\pi)+k^2/(8h)) \) is the radiation damping term for a metamaterial embedded in the PPW structure. In expression~\eqref{eq: rr_correction}, \(\mathbf{A}_n\) denotes the ``effective'' polarizability, which contains the radiation damping and satisfies the power conservation constraint, while $\tilde{\mathbf{A}}_n$ denotes the intrinsic polarizability, which is obtained from a quasi-static model \cite[Table~(12.1)]{collin1990field}, and it only depends on the metamaterial's characteristics, i.e., geometry and impedance loading. In this paper, we consider each metamaterial to be reconfigurable and symmetric w.r.t. the axes of the PPW, so the cross-polarization terms of the polarizability can be considered zero, thus, making \(\mathbf{A}_n\) diagonal. The most realistic choice to enforce this symmetry for reconfigurable elements is that the tunable variables in the \(x\) and \(y\) components are identical, e.g., two varactor diodes with common bias. Then, tailoring the Lorentzian phase mapping \cite{davidsmith2017}, by accounting for the radiation damping and the correct sign conventions, we write the polarizability matrix of each \(n\)-th element as follows:
\begin{equation}\label{eq:lorentzian_mapping_appendix}
\mathbf{A}_n
=
-\frac{1}{2C}\big(\jmath+e^{\jmath\varphi_n}\big)\mathbf{I}_2,
\end{equation}
for some tunable phase \(\varphi_n\in[0,2\pi]\). This mapping satisfies \({\rm Im}\{\mathbf{A}_n^{-1}\}=C\) and \({\rm Im}\{\mathbf{A}_n\}\leq \mathbf{0}\), which are the necessary conditions for lossless and passive dipoles \cite{Gavriilidis2026WaveguideMetasurface}.

\subsection{Near-Field Free-Space Channel}

To compute the electric field in free space, we use \(\mathbf{e}_{{\rm sc},n}(\mathbf{s})\triangleq\jmath \omega \mu_0 [\nabla \!\times \!\tilde{\mathbf{G}}_{\rm FS}(\mathbf{s}-\mathbf{r}_n)]\mathbf{m}_n\), where \(\tilde{\mathbf{G}}_{\rm FS}\in \mathbb{C}^{3 \times 3}\) is the dyadic Green function as given in \cite[eq.~(8.61)]{Novotny_Hecht_2006}, but doubled due to the dipole image created by the top plate of the waveguide. Then, the scattered electric field is expressed in spherical coordinates, and we consider the longitudinal component \(r_n\) per $n$-th element to be negligible compared to the azimuth, \(\phi_n\), and elevation, \(\theta_n\), components. Consequently, the electric field at an observation point \(\mathbf{s}\) due to the dipole located at \(\mathbf{r}_n\), has the following two components:\vspace{-0.1cm}
\begin{align}\label{eq: electric_field_due_to_nth_dipole}
\!\!e_{{\rm sc},n}^{\theta}(\mathbf{s})\!
&=\!
\frac{\eta k^{2}e^{-\jmath kR_{n}}}{2\pi  R_{n}}\!
\big(\!m_{n}^{x}\sin(\!\phi_{n}\!) \!-\! m_{n}^{y}\cos(\!\phi_{n}\!)\!\big),\\[2mm]
\!\! e_{{\rm sc},n}^{\phi}(\mathbf{s})
\!&=\!
\frac{\eta k^{2}e^{-\jmath kR_{n}}}{2\pi R_{n}}
\!\big(\!m_{n}^{x}\cos(\!\phi_{n}\!) \!+ \!m_{n}^{y}\sin(\!\phi_{n}\!)\!\big)\!\cos(\!\theta_{n}\!),\nonumber\vspace{-0.1cm}
\end{align}
where \(\eta\) is the free-space impedance, \(R_{n}\) is the radial distance between \(\mathbf{s}\) and \(\mathbf{r}_n\), and \(\theta_{n} ={\rm acos}\left(({s_{z} - r_{{z}_n}})/{R_{n}}\right)\) as well as \(\phi_{n} = {\rm atan}\left( ({s_{y}- r_{y_n}})/({s_{x}- r_{x_n}})\right)\) are the elevation and azimuth angles between them. 
Importantly, the elevation and azimuth angles between each \(n\)-th dipole and the observation point, $\theta_{n}$ and $\phi_{n}$, depend on the specific dipole location
$\mathbf{r}_n$, and thus the field components $e_{{\rm sc},n}^{\theta}(\mathbf{s})$ and $e_{{\rm sc},n}^{\phi}(\mathbf{s})$ in \eqref{eq: electric_field_due_to_nth_dipole} are expressed in
dipole-dependent local spherical bases. As a result, these components do not correspond to a common polarization basis and cannot be summed \cite{gavriilidis2026nearfield}. 

To compute the total field, each \(n\)-th contribution is projected onto a common spherical basis defined at the observation point \(\mathbf{s}\) w.r.t. the TX center. Let $(\theta_s,\phi_s)$ denote the angles of $\mathbf{s}$ w.r.t. the TX center, and let $\hat{\boldsymbol{\theta}}_s$, $\hat{\boldsymbol{\phi}}_s$ represent the associated spherical unit vectors. Similarly, $\hat{\boldsymbol{\theta}}_{s,n}$ and $\hat{\boldsymbol{\phi}}_{s,n}$ correspond to the direction $\mathbf{s}-\mathbf{r}_n$ \cite{gavriilidis2026nearfield}. Then, we can define the $2\times2$ projection matrix
\(\mathbf T_{n\rightarrow s}
\triangleq [\hat{\boldsymbol{\theta}}_s^{\mathrm T}\hat{\boldsymbol{\theta}}_{s,n}, \, 
\hat{\boldsymbol{\theta}}_s^{\mathrm T}\hat{\boldsymbol{\phi}}_{s,n};\,
\hat{\boldsymbol{\phi}}_s^{\mathrm T}\hat{\boldsymbol{\theta}}_{s,n},\, 
\hat{\boldsymbol{\phi}}_s^{\mathrm T}\hat{\boldsymbol{\phi}}_{s,n}]\),
which maps the components of the \(n\)-th dipole, \(e_{{\rm sc},n}^{\phi}(\mathbf{s})\) and \(e_{{\rm sc},n}^{\theta}(\mathbf{s})\), from their local basis into the common basis. To express all dipole contributions in the common spherical basis, we first introduce the local focusing vectors $\mathbf a^{\theta}(\mathbf s)$ and
$\mathbf a^{\phi}(\mathbf s)$, which collect the dipole-dependent responses prior to basis unification, and whose $n$-th block, $\forall n=1,\ldots,N$, of size $2\times1$ is given as follows:
\vspace{-0.2cm}
\begin{align}
  \left[[\mathbf a^{\theta}]_{2n-1},
[\mathbf a^{\theta}]_{2n}\right]^{\rm T}
& =
\frac{e^{-jkR_n}}{R_n}
[\sin(\phi_n),
-\cos(\phi_n)]^{\rm T},\\  
\left[[\mathbf a^{\phi}]_{2n-1},
[\mathbf a^{\phi}]_{2n}
\right]^{\rm T}
&=\frac{e^{-jkR_n}\cos(\theta_n)}{R_n}
[\cos(\phi_n),
\sin(\phi_n)
]^{\rm T}.\nonumber\vspace{-0.2cm}
\end{align}
Consequently, to account for the basis mismatch, the focusing vectors are transformed
element-wise through $\mathbf T_{n\rightarrow s}$, yielding the
projected (common-basis) focusing vectors
$\tilde{\mathbf a}^{\theta}(\mathbf s)\in\mathbb C^{2N\times1}$ and
$\tilde{\mathbf a}^{\phi}(\mathbf s)\in\mathbb C^{2N\times1}$, defined
for $n=1,\ldots,2N$ as:\vspace{-0.1cm}
\begin{equation}\label{eq:projected_focusing_vectors}
 \left[   [\tilde{\mathbf a}^{\theta}(\mathbf s)]_{n},
 [\tilde{\mathbf a}^{\phi}(\mathbf s)]_{n} \right]^{\rm T}  
=\mathbf{T}_{n\rightarrow s}\left[[\mathbf a^{\theta}(\mathbf s)]_{n},
[\mathbf a^{\phi}(\mathbf s)]_{n}\right]^{\rm T}.\vspace{-0.1cm}
\end{equation}
Then, the scattered field components at $\mathbf{s}$ are given as follows:\vspace{-0.1cm}
\begin{equation}\label{eq: total electric field_projected}
[e_{\rm sc}^{\theta}(\mathbf{s}),
e_{\rm sc}^{\phi}(\mathbf{s})]=
{\eta k^{2}}/({2\pi})\,
\left[\tilde{\mathbf a}^{\theta}(\mathbf s)^{\mathrm T}\mathbf m,\tilde{\mathbf a}^{\phi}(\mathbf s)^{\mathrm T}\mathbf m \right].\vspace{-0.1cm}
\end{equation}
Using \eqref{eq: total electric field_projected}, we define the dual-polarized channel between the TX and \(\mathbf{s}\) as
\(\mathbf{H}(\mathbf{s}) \in \mathbb{C}^{2\times 2N}\), given by: \([\mathbf H(\mathbf{s})]_{1,:}
\triangleq
{\eta k^{2}}/({2\pi})\tilde{\mathbf a}^{\theta}(\mathbf s)^{\mathrm T}\) and \([\mathbf H(\mathbf{s})]_{2,:}
\triangleq
{\eta k^{2}}/({2\pi})\,
\tilde{\mathbf a}^{\phi}(\mathbf s)^{\mathrm T}.\) Then, the electric signal \(\mathbf{y}\in \mathbb{C}^{2 \times 1}\) received at \(\mathbf{s}\) can be expressed as: 
\vspace{-0.15 cm}
\begin{equation}\label{eq: Received Signal}
    \mathbf{y} \triangleq \mathbf{H}(\mathbf{s}) \left(\bar{\mathbf{A}}^{-1} - \bar{\mathbf{G}}\right)^{-1}\mathbf{h}_f I.
\end{equation}

\section{Near-Field Gain Scaling and Beam Depth}
In this section, we characterize the NF behavior of the considered 2D waveguide-fed metasurface antenna. Specifically, we first study how the beam focusing gain scales with the aperture radius \(D\), and then analyze the beam depth, i.e., the ability of the system to provide spatial selectivity along the radial direction. To obtain tractable and angle-independent insights, we focus on an observation point located along the normal direction of the metasurface, defined as \(\mathbf{s}(R) \triangleq [0,0,R]^{\mathrm T}\). This direction typically corresponds to the maximum directivity of aperture antennas, and thus provides a representative (and upper-bound) characterization of NF performance. After projection to the common spherical basis, the scattered fields at \(\mathbf{s}(R)\) become as follows:
\begin{equation}\label{eq: theta component sR}
e_{\rm sc}^{\theta}(\mathbf{s}(R))
=
-\frac{\eta k^2}{2\pi}
\sum_{n=1}^N
\frac{R}{R_n^2(R)} e^{-jkR_n(R)} m_n^y,
\end{equation}
\begin{equation}\label{eq: phi component sR}
e_{\rm sc}^{\phi}(\mathbf{s}(R))
=
\frac{\eta k^2}{2\pi}
\sum_{n=1}^N
\frac{R}{R_n^2(R)} e^{-jkR_n(R)} m_n^x,
\end{equation}
where \(R_n(R)\triangleq \sqrt{R^2+\rho_n^2}\) and \(\rho_n\triangleq \sqrt{x_n^2+y_n^2}.\)
At this stage, to derive asymptotic relations and scaling laws for the beam focusing performance of the PPW architecture, we neglect mutual coupling, i.e., \(\bar{\mathbf{G}}\approx\mathbf{0}\); in the upcoming Section~\ref{sec: Simulations}, the resulting expressions will be validated against the full model that accounts for mutual coupling. We design the metasurface to maximize the $\theta$-polarized field, which depends only on $m_n^y$. Using the Lorentzian mapping from \eqref{eq:lorentzian_mapping_appendix},
we select for each $n$-th element:
\begin{equation}\label{eq: varphi_n solution}
\varphi_n = kR_n(R) - \angle(h_{0,n}^y),
\end{equation}
so that the coherent phase of $m_n^y = A^{yy}_n h_{0,n}^y$ aligns with the propagation phase of the guided field. Substituting \(\varphi_n\) from \eqref{eq: varphi_n solution}, yields \(m_n^y =-\frac{1}{2C}\left(\jmath h_{0,n}^y + |h_{0,n}^y| e^{jkR_n(R)}\right)\). The first term does not satisfy the focusing phase condition and is therefore asymptotically negligible in the aperture summation. To this end, we use the approximation \(m_n^y \approx -\frac{1}{2C}|h_{0,n}^y| e^{jkR_n(R)}.\) Substituting this into the field expression \eqref{eq: theta component sR} and expanding the guided magnetic field from \eqref{eq: Excitation Field}, we obtain:
\begin{equation}\label{eq: theta component coherent sum}
e_{\rm sc}^{\theta}(\mathbf{s}(R))
\!\approx\!
\frac{\eta k^3 |I|}{16\pi C}
\sum_{n=1}^N
\frac{R}{R_n^2(R)}
\left|H_1^{(2)}(k\rho_n)\right||\cos(\psi_n)|.
\end{equation}

We now examine the $\phi$-polarized field under the considered phase design 
\(
m_n^x = -\frac{1}{2C}\left(\jmath h_{0,n}^x + e^{\jmath\varphi_n} h_{0,n}^x\right).
\)
Then, the coherent component becomes \(e^{\jmath\varphi_n} h_{0,n}^x
=
|h_{0,n}^x| e^{jkR_n(R)} \operatorname{sgn}(-\tan(\psi_n)).\)
Due to the radially propagating wave, it is convenient to think of the metasurface as a dense disk of metamaterials centered at the source. Hence, decomposing the aperture into quadrants centered at the feed, the sign term $\operatorname{sgn}(\!\text{-}\!\tan(\psi_n)\!)$ alternates across opposite quadrants. Since elements at equal $\rho_n$ share identical \(R_n(R)\), the contributions to the $\phi$-polarized field cancel pairwise, yielding \(e_{\rm sc}^{\phi}(\mathbf{s}(R)) \!\!\approx \!\! 0.\) Consequently, maximizing the $\theta$-polarized component results in an asymptotic nulling of the $\phi$-polarized field. The reverse holds if the phase is designed to maximize \(e^{\phi}(\mathbf{s}(R))\).

We proceed by approximating the sum in \eqref{eq: theta component coherent sum} by its continuum integral representation, by modeling the aperture as a disk of radius \(D\) populated by \(N=\nu\pi D^2\) metamaterials, where \(\nu\) denotes the surface density, yielding:
\begin{align}\label{eq:theta_field_disk_integral}
e_{\rm sc}^{\theta}(\mathbf{s}(R))
\approx &
\frac{\eta k^3 |I|}{16\pi C}\,
\nu \\
& \int_{0}^{2\pi}\int_{0}^{D}
\frac{R}{R^2+\rho^2}
\left|H_1^{(2)}(k\rho)\right||\cos(\psi)|
\rho\,d\rho\,d\psi,\nonumber
\end{align}
where we used \(R_n^2(R)=R^2+\rho^2\), with \(\rho\) and \(\psi\) denoting the polar coordinates on the disk. Since \(\int_{0}^{2\pi} |\cos(\psi)|\,d\psi = 4\) and using the large-argument asymptotic behavior \(|H_1^{(2)}(k\rho)| \sim \sqrt{\frac{2}{\pi k\rho}}\) to substitute into \eqref{eq:theta_field_disk_integral}, yields the approximation:
\begin{equation}\label{eq:theta_field_scaling_integral}
e_{\rm sc}^{\theta}(\mathbf{s}(R))
\approx
\frac{\eta k^3 |I|\,\nu}{4\pi C}
\sqrt{\frac{2}{\pi k}}
\int_{0}^{D}
\frac{R}{R^2+\rho^2}
\rho^{1/2}\,d\rho.
\end{equation}
To obtain a closed-form scaling law, we consider the regime \(D\ll R\), for which: \(\frac{R}{R^2+\rho^2}\approx \frac{1}{R}.\) Then, \eqref{eq:theta_field_scaling_integral} reduces to:
\begin{equation}\label{eq: theta solved integral}
e_{\rm sc}^{\theta}(\mathbf{s}(R))
\approx
\frac{\eta k^3 |I|\,\nu}{6\pi CR }
\sqrt{\frac{2}{\pi k}} { D^{3/2}}.
\end{equation}
Hence, the \(\theta\)-polarized field amplitude scales as
\(e_{\rm sc}^{\theta}(\mathbf{s}(R)) \propto \nu D^{3/2}\), and, since the number of elements is \(N=\nu \pi D^2\), we equivalently obtain that the squared electric field scales, when assuming \(\nu\) is fixed, as follows:
\begin{equation}\label{eq:Etheta_squared_scaling_N}
|e_{\rm sc}^{\theta}(\mathbf{s}(R))|^2 \propto N^{3/2}.
\end{equation}
The scaling law in \eqref{eq:Etheta_squared_scaling_N} holds in the regime where \(D\ll R\). As \(D\to\infty\), the integral in \eqref{eq:theta_field_scaling_integral} converges to \(\pi \sqrt{0.5R}\), indicating saturation of the field magnitude with increasing aperture size. Notably, the \(N^{3/2}\) scaling arises from coherent combining across the aperture, and also applies to FF beamforming under appropriate phase alignment.

It is important to emphasize that the scaling law \(|e_{\rm sc}^{\theta}(\mathbf{s}(R))|^2 \propto N^{3/2}\) is obtained under the assumption of fixed feed current \(I\). However, this is not the appropriate normalization for comparing how the beam focusing performance evolves with the number of metamaterial elements. Specifically, changing \(N\), modifies the equivalent impedance of the PPW structure seen by the feed, and therefore a fixed current excitation does not correspond to a fixed accepted power. To circumvent this issue, the field scaling should be normalized by the supplied power to the metamaterials. To this end, using the Poynting theorem for dipoles \cite[eq.~(8.74)]{Novotny_Hecht_2006}, we can express the supplied power as follows:
\begin{equation}\label{eq: supplied power}
P_{\rm sup}
\triangleq
\frac{1}{2}\Re\!\left\{
\jmath\omega \mu_0
\sum_{n=1}^N
\mathbf m_n^{\mathrm T}\mathbf h_{\text{loc},n}^*
\right\}.
\end{equation}
Under the approximation with no coupling, \(\mathbf h_{\text{loc},n} = \mathbf{h}_{0,n}\) and \(\mathbf m_n=\mathbf A_n\mathbf h_{0,n}\). Moreover, taking into account the diagonal form of \(\mathbf{A}_n\), i.e., \(\mathbf A_n={\rm diag}(A_n^{xx},A_n^{yy})\),\eqref{eq: supplied power} reduces to:
\begin{equation}
P_{\rm sup}
=
\frac{\omega \mu_0}{2}
\sum_{n=1}^N
\Big(
-\Im\{A_n^{xx}\}|h_{0,n}^x|^2
-
\Im\{A_n^{yy}\}|h_{0,n}^y|^2
\Big).
\end{equation}
Next, a per the symmetric polarizability model in \eqref{eq:lorentzian_mapping_appendix}, \(A_n^{xx}=A_n^{yy}=A_n\), and therefore:
\begin{equation}
P_{\rm sup}
=
\frac{\omega \mu_0 k^2|I|^2}{32}
\sum_{n=1}^N
\left|H_1^{(2)}(k\rho_n)\right|^2
\big(-\Im\{A_n\}\big),
\end{equation}
where we used \(|h_{0,n}^x|^2+|h_{0,n}^y|^2=\frac{k^2|I|^2}{16}|H_1^{(2)}(k\rho_n)|^2\). Furthermore, approximating the dense metasurface by a circular aperture of radius \(D\), using \(|H_1^{(2)}(k\rho)|^2\sim 2/(\pi k\rho)\), and substituting \(-\Im\{A_n\}=\frac{1+\sin(\varphi_n)}{2C},\) we can approximate as:
\begin{equation}
P_{\rm sup}
\approx
\frac{\omega\mu_0 k|I|^2\nu}{32\pi C}
\int_0^{2\pi}\int_0^D
\big(1+\sin\!\left(\varphi(\rho,\psi)\right)\big)\,d\rho\,d\psi.
\end{equation}
The term containing \(\sin\!\left(\varphi(\rho,\psi)\right)\) is oscillatory, as it contains both the focusing phase \(kR_n\) and the phase of the local excitation. Hence, as a function of \(\rho\) and \(\psi\), it varies rapidly over the aperture and does not admit coherent accumulation in the integral above. Therefore, its contribution is asymptotically negligible compared to that of the constant term. By retaining only the dominant contribution, yields \(P_{\rm sup}= \omega \mu_0 k I^2 \nu D /(16 C)\). Then,  defining the gain at \(\mathbf{s}(R)\) as \(G(R) \!\triangleq\! |e_{\rm sc}(\mathbf{s}(R))|^2/P_{\rm sup}\), we reach the following~expression:
\begin{equation}\label{eq: analytic gain}
G(R) = \frac{8 \eta k^3 \nu}{9 \pi^3 C R^2} D^2.
\end{equation}
Hence, the gain scales with the square of the radius \(D\), and linearly with \(N\), since \(N=\nu\pi D^2\), despite the non-uniform excitation across the aperture. This linear scaling is physically intuitive: although the excitation decays with distance from the feed, the circumference of each concentric ring increases proportionally with radius, so the larger number of elements at outer radii compensates for their weaker excitation, yielding an overall gain that scales linearly with \(N\).


\subsection{Characterization of the Beam Depth}
We now characterize the beam-depth function by considering a mismatch in the focal distance. Specifically, the metasurface is designed to focus at \(\mathbf{s}(R+\Delta\! R)\), while the actual observation point remains \(\mathbf{s}(R)\). Under the same no-coupling approximation as above, this mismatch only alters the phase of each dipole contribution, while the excitation-induced amplitude taper remains unchanged. Therefore, the \(\theta\)-polarized field at \(\mathbf{s}(R)\) can be approximated as follows:
\begin{align}
& e_{\rm sc}^{\theta}(\mathbf{s}(R);R+\Delta \!R)
= 
\frac{\eta k^3 |I|}{16\pi C} \nonumber\\
&\times\sum_{n=1}^N
\frac{R}{R_n^2(R)}
\left|H_1^{(2)}(k\rho_n)\right||\cos\psi_n|
e^{-jk\left(R_n(R+\Delta\! R)-R_n(R)\right)}.\nonumber
\end{align}
Accordingly, the normalized beam-depth gain is defined as the ratio between the gain at \(\mathbf{s}(R)\) when focusing at \(\mathbf{s}(R+\Delta R)\) and the gain when correctly focusing at \(\mathbf{s}(R)\), yielding:
\begin{align}\label{eq:beam_depth_discrete}
&\bar{G}(R,\Delta R)
\triangleq\\
&\frac{
\left|
\displaystyle\sum_{n=1}^N
\frac{R}{R_n^2(R)}
\left|H_1^{(2)}(k\rho_n)\right||\cos\psi_n|\,
e^{-jk\left(R_n(R+\Delta\! R)-R_n(R)\right)}
\right|^2
}{
\left|
\displaystyle\sum_{n=1}^N
\frac{R}{R_n^2(R)}
\left|H_1^{(2)}(k\rho_n)\right||\cos\psi_n|
\right|^2
}\nonumber.
\end{align}

Following the prior analysis, we approximate both terms as an integral over a dense circular aperture of radius \(D\). Then, the integral over \(\psi\) and any constant scalings cancel out, since they are identical in both terms. Furthermore, we use the large argument form of the Hankel function, and use \(R/R^2_n(R)\approx 1/R\), which cancels out due to the division. After these simplifications, \(\bar{G}(R,\Delta R)\) can be approximated as:
\begin{equation}\label{eq:beam_depth_radial}
\bar G(R,\Delta R)
\approx
\frac{
\left|
\int_0^D
\rho^{1/2}
e^{-jk\left(\sqrt{(R+\Delta R)^2+\rho^2}-\sqrt{R^2+\rho^2}\right)} d\rho
\right|^2
}{
\left|
\int_0^D
\rho^{1/2}\,d\rho
\right|^2
}.
\end{equation}
Using the Fresnel approximation
\(
\sqrt{(R+\Delta\! R)^2+\rho^2}-\sqrt{R^2+\rho^2}
\approx
\Delta R-\frac{\Delta\! R}{2R(R+\Delta\! R)}\rho^2,
\)
omitting the global phase factor \(e^{-jk\Delta \!R}\), which vanishes after taking the modulus, and computing the integral in the denominator, \eqref{eq:beam_depth_radial} becomes:
\begin{equation}\label{eq: beam-depth gain approximation}
\bar G(R,\Delta\! R)
\approx
\frac{9}{4} D^{-3}{
\bigg|
\underbrace{\int_0^D
\rho^{1/2}e^{\jmath\alpha(\Delta \! R)/D^2 \rho^2}\,d\rho}_{\triangleq\mathcal{I}(\alpha(\Delta \! R))}
\bigg|^2
},
\end{equation}
where \(\alpha(\Delta\! R) \triangleq \frac{D^2 k\Delta \!R}{2R(R+\Delta R)}.\) This expression can be further simplified by introducing the normalized variable \(t \triangleq \frac{\rho^2}{D^2},
\,d\rho = \frac{D}{2}t^{-1/2}dt.\)
Substituting into the integral yields:
\begin{equation}\label{eq: nominator integral}
\mathcal I(\alpha(\Delta \!R))
=
\frac{D^{3/2}}{2}
\int_0^1
t^{-1/4}e^{\jmath\alpha(\Delta \!R) t}\,dt.
\end{equation}
Hence, substituting \eqref{eq: nominator integral} to \eqref{eq: beam-depth gain approximation} and changing the argument of \(\bar{G}(\cdot)\) from \((R,\Delta\! R)\) to \(\alpha(\Delta\! R)\), we reach the expression:
\begin{equation}\label{eq: analytic beam depth gain}
\bar{G}(\alpha(\Delta\! R)) =
\frac{9}{16}\left|\int_0^1
 t^{-1/4}e^{\jmath\alpha(\Delta\! R) t}\,dt\right|^2\!.
\end{equation}
Note that this expression can be alternatively expressed in terms of the lower incomplete gamma function, if a closed-form representation is preferred.
Moreover, it explicitly shows that the beam-depth function depends only on \(\alpha(\Delta R)\). Hence, the beam-depth limits are obtained by solving \(\bar{G}(\alpha(\Delta \!R))=\kappa\), where \(\kappa\in [0,1]\) denotes the fraction of the maximum gain attained under a radial mismatch \(\Delta\! R\). The beam depth is defined as the range \(\Delta\! R \in[\Delta \!R^{-}\!(\kappa),\Delta\! R^{+}\!(\kappa)]\) within which \(\bar{G}(\alpha(\Delta\! R))\geq \kappa\). The function \(\bar{G}(\alpha)\) is observed to be decreasing (up to mild oscillations), as it corresponds to the squared magnitude of a Fourier-type transform of a decreasing function. Let \(\alpha_{\kappa}\) denote the smallest positive solution of \(\bar{G}(\alpha_\kappa)=\kappa\); the interval bounds are then given by:
\begin{equation}\label{eq: beam-depth limits}
    \Delta\!R ^{+}\!(\kappa) = \frac{2 R^2 \alpha_\kappa}{D^2 k - 2 R \alpha_\kappa},\, \Delta\!R ^{-}\!(\kappa) = \frac{-2 R^2 \alpha_\kappa}{D^2 k + 2 R \alpha_\kappa}.
\end{equation}
It can be seen that the limit \(\Delta\!R^{+}\!(\kappa)\) tends to infinity as \(R \to R_{{\rm lim},\kappa}\triangleq D^2k/(2 \alpha_{\kappa})\). This point actually marks a transition to FF-like behavior for the gain level \(\kappa\), in the sense that, for \(R\geq R_{{\rm lim},\kappa}\), the condition \(\bar{G}(\alpha(\Delta\! R))\geq \kappa\) holds for arbitrarily large positive radial offsets \(\Delta\! R\). Consequently, the beam becomes insensitive to positive displacements along the radial direction, and the orthogonality in \(R\) is effectively lost, yielding FF-like behavior.

\begin{figure}
    \centering
    \begin{subfigure}{\linewidth}
        \centering
        \includegraphics[width=\linewidth]{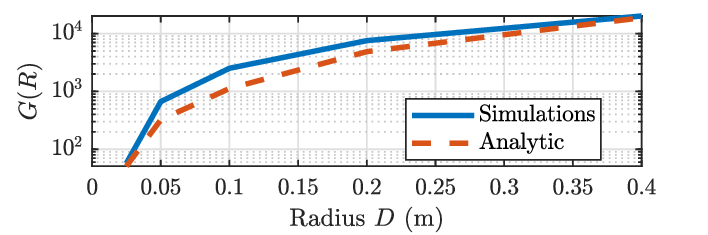}
        \caption{}
        \label{fig: gain vs D}
    \end{subfigure}
        
    \begin{subfigure}{\linewidth}
        \centering
        \includegraphics[width=\linewidth]{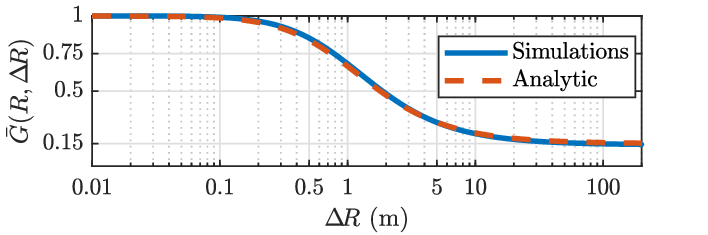}
        \caption{}
        \label{fig: beam depth gain vs DR}
    \end{subfigure}
    
    \caption{(a) The gain function, defined as \(G(R)=\frac{|e_{\rm sc}(\mathbf{s}(R))|^2}{P_{\rm sup}}\), for increasing aperture radius \(D\); comparison between the analytic function in \eqref{eq: analytic gain} and the simulated \(G(R)\). 
    (b) The normalized beam-depth gain \(\bar{G}(R,\Delta \! R)\) versus the analytic function given at \eqref{eq: analytic beam depth gain} for increasing positive \(\Delta\! R\) values.}
    \label{fig: combined gain}
    \vspace{-0.6 cm}
\end{figure}

\vspace{-0.2 cm}
\section{Simulation Results and Conclusion} \label{sec: Simulations}
In this section, we validate the analytic gain and beam-depth expressions against simulations based on the full electromagnetic model that accounts for mutual coupling. Although coupling was neglected in the derivations to enable closed-form analysis, the results demonstrate that the predicted scaling laws and beam-depth limits remain accurate under the complete model. In particular, the dipole moments in simulations were computed as \(\mathbf{m} = (\mathbf{A}^{-1}-\bar{\mathbf{G}})^{-1}\mathbf{h}_{\rm f}I\), in contrast to the no-coupling approximation \(\mathbf{m} = \mathbf{A}\mathbf{h}_{\rm f}I\) used in the analysis; these are referred to as “Simulations” and “Analytic” results, respectively. Moreover, the closed-form phase configuration in \eqref{eq: varphi_n solution} is no longer optimal in the presence of coupling. Therefore, in simulations, the phase configuration \(\boldsymbol{\varphi}\) was optimized by maximizing \(|e^{\theta}(\mathbf{s}(R))|^2\) using Riemannian Manifold Optimization (RMO) on the complex circle manifold with a conjugate gradient algorithm \cite[Alg.~1]{Alexandropoulos2023CounterActingEaves}. Unless otherwise stated, the observation point was located at \(R=1\) m from the aperture center. The system operated at \(f=10\) GHz with waveguide height \(h=2\) mm. The metasurface elements were arranged in concentric rings around a central feed, with both radial and angular spacings between~adjacent~elements~set~to~\(\lambda/2\).

Figure~\ref{fig: gain vs D} illustrates the scaling performance of the PPW-fed metasurface w.r.t. its radius $D$. In particular, we compared the analytic gain function from \eqref{eq: analytic gain} to the simulated gain achieved using the coupled modeling. For each aperture radius \(D\), the phase configuration \(\boldsymbol{\varphi}\) was optimized via RMO. The results show that the analytic scaling closely matches the simulated trend, despite the different modeling assumptions, confirming the robustness of the derived law.  For small radii, where mutual coupling is weak, the agreement is particularly tight. As \(D\) increases, the two curves remain close and eventually converge again for large apertures. This behavior is consistent with the asymptotic nature of the analysis, which becomes more accurate as \(N\) increases and incoherent contributions vanish. For reference, we note that \(D=0.05\) m and \(0.4\) m correspond to \(N=38\) and \(2206\) elements, respectively.

Figure~\ref{fig: beam depth gain vs DR} validates the normalized beam-depth gain analysis by comparing the simulated \(\bar{G}(R,\Delta \! R)\) with the derived analytic expression in \eqref{eq: analytic beam depth gain}. As shown, the two curves exhibit very close agreement over \(\Delta\! R\), confirming the accuracy of the derived model. In this setup, the aperture radius is \(D=0.2\) m, for which \(R_{\lim,0.15}=1\) m. Since the observation point is fixed at \(R=1\) m, we are effectively at the limit \(R\geq R_{\lim,0.15}\). According to the analysis, this implies that, for any positive radial offset \(\Delta R>0\), the normalized beam-depth gain remains above \(0.15\). This behavior is verified in the figure, where it converges to \(0.15\) as \(\Delta\! R\) increases.

Concluding, although the presented scaling and beam-depth analysis neglected mutual coupling, the resulting analytic laws extend to the full system behavior. This is because the scaling is fundamentally governed by aperture-level coherent accumulation, while mutual coupling modifies the local fields and thus the optimal phase configuration. When properly accounted for, these interactions can be exploited to enhance the radiated field in the desired direction, yielding gains that can even exceed the analytic scaling law, as it was shown in Fig.~\ref{fig: gain vs D}.




\bibliographystyle{IEEEtran}
\vspace{-0.3 cm}
\bibliography{references}
\vspace{-0.4 cm}
\end{document}